\documentclass[prb,preprint,unsortedaddress]{revtex4-1}
\usepackage{amsmath}  % needed for \tfrac, \bmatrix, etc.
\usepackage{amsfonts} % needed for bold Greek, Fraktur, and blackboard bold
\usepackage{graphicx} % needed for figures

\begin{document}

% Be sure to use the \title, \author, \affiliation, and \abstract macros
% to format your title page.  Don't use lower-level macros to  manually
% adjust the fonts and centering.

\title{Axiomatic quantum mechanics: Necessity and benefits for the physics studies}
% In a long title you can use \\ to force a line break at a certain location.

\author{Jasmina Jekni\' c-Dugi\' c}
\affiliation{University of Ni\v s, Faculty of Science and Mathematics, 18000 Ni\v s, Serbia}

\author{Momir Arsenijevi\' c}
\affiliation{University of Kragujevac, Faculty of Science, 34000 Kragujevac, Serbia}
% Please provide a full mailing address here.

\author{Miroljub Dugi\' c}
\email{mdugic18@sbb.rs} % optional
\affiliation{University of Kragujevac, Faculty of Science, 34000 Kragujevac, Serbia}

% See the REVTeX documentation for more examples of author and affiliation lists.

\date{\today}

\begin{abstract}

The ongoing progress in quantum theory emphasizes the crucial role of the very basic principles of quantum theory. However, this is not properly followed in teaching quantum mechanics on the graduate and undergraduate levels of physics studies. The existing textbooks typically avoid the axiomatic presentation of the theory. We emphasize usefulness of the systematic, axiomatic approach to the basics of quantum theory as well as its importance in the light of the modern scientific-research context.

\end{abstract}
% AJP requires an abstract for all regular article submissions.
% Abstracts are optional for submissions to the "Notes and Discussions" section.

\maketitle % title page is now complete

{\bf 1. Introduction}

\bigskip

Even a brief search on the Internet reveals a significant delay in the textbook presentation of the basics of quantum mechanics relative to the recent and ongoing, considerable scientific and technological progress that is {\it based on the basic principles of quantum mechanics}, e.g. [1-3]. New reading and reconsideration of the basic postulates of quantum mechanics generate and accumulate the knowledge that should be properly presented on the graduate or undergraduate physics studies.

	According to the data available on the Internet, by far the most of the most-used graduate/undergraduate-level textbooks poorly present the axiomatic and methodological character of quantum mechanics. Plenty of the textbooks clearly serve the purposes of application e.g. in the atomic or molecular or condensed matter physics, with an emphasis on the {\it special} methods and models at the expense of the clearly presented {\it universal methods and formalism}. This situation seems fairly presented by the following quote taken over from an Internet discussion on ''What are some good resources for learning quantum mechanics?'' [our bold-face emphasis]:

\noindent ''{\it Quantum mechanics is actually sort of a pain when it comes to textbooks - you can read one textbook and still come out clueless when reading another - in other words - QM books are often so different that you may need to read several books to really understand the subject [.] Anyways, the standard book that most undergrads use is [...] With that said, it barely uses the convenient Dirac notation at all} {\bf (as it assumes that students are ignorant of basic linear algebra)}, {\it and many complain about its lack of rigor.}''

	In effect, the students are left to memorizing the contents, instead of the relatively simple deductive thinking. For example, the extensive use of the historical ad hoc rules for the ''quantum jumps'' in atomic physics neither opens a door to insight in the modern approach to the topic, e.g. [5, 6], nor does it provide a link to the alternative modern approaches within the open system’s theory [6, 7], in which e.g. the hydrogen atom is {\it not} regarded an isolated but an {\it open} quantum system [2, 6-8]. It is a matter of fact that Dirac's completion [4] and von Neumann's extension [9] of the unification of the ad-hoc schemes of the wave- and the matrix-quantum-theories are too often ignored by the authors and teachers on the graduate and undergraduate level. According to Zurek's [10]:

\noindent ''{\it Quantum mechanics has been to date, by and large, presented in a manner that reflects its historical development. That is, Bohr's planetary model of the atom is still often the point of departure, Hamilton-Jacobi equations are used to 'derive' the Schr\" odinger equation, and an oversimplified version of the quantum-classical relationship (attributed to Bohr, but generally not doing justice to his much more sophisticated views) with the correspondence principle, kinship of commutators and Poisson brackets, the Ehrenfest theorem, some version of the Copenhagen interpretation, and other evidence that quantum theory is really not all that different from classical--especially when systems of interest become macroscopic, and all one cares about are averages--is presented.}''

	What might had made the excellent textbooks e.g. of Messiah [11] and Landau \& Lifshitz [12], even of Cohen Tanoudji et al [13], practically superfluous--according to the numerous informal student polls that can be found on the Internet? [The only reason we do not provide the links is to avoid the possible abuse or to be offensive due to someone's opinion.]

	We suspect the at least double origin for this state of affair. On the one hand, there was a famous say and wisdom of ''shut up and calculate.'' In this context, quantum mechanics was regarded merely a ''cookbook'' for diverse applications--the ''cookbook'' might be enhanced but hardly of any further and deeper scientific interest. On the other hand, the full mathematical foundations [10] are rather demanding and time-consuming for the starters. Therefore some balance is required for the introductory course in quantum mechanics. Finding this balance is not an easy task; neither the unique output can be expected, e.g. [14-17]. Nevertheless, and this is the main point of this paper, {\it the axiomatic and methodological character of quantum theory can easily be appreciated in a textbook}, with the significant benefits for the students. Here we comply with Diosi's [6]:

\noindent ''{\it We should build as much as possible on standard knowledge, using standard concepts, equations, terminology.}''

\noindent that emphasizes a need for a proper preparation of students before studying the quantum-mechanical theory.

	Bearing in mind the unprecedented success of the {\it direct} use of the very foundations of quantum theory, e.g. [1-3], the time might had probably come to reaffirm the axiomatic theory [1, 4, 10] that was carefully (also prepared for the advanced studies) presented in some classic textbooks [11-13] on quantum mechanics.

	Leaving aside the scientific strictness as well as a dispute among the foundationally-oriented and the application-oriented authors, we emphasize the following benefits for the students familiar with the axiomatic quantum-mechanical theory. First of all, deductive thinking is clear and relatively easy to follow. It leads to the scientific ideal: theoretically reliable predictions. Second, the concepts are given in a clear form
thus reducing, and often removing, the room for "mystical"  contents of the theory. Third, clearly defined concepts open the room for {\it critical} thinking--the weak points in the theory can be easily detected and
this may serve as a starting point for both the students independent thinking as well as a first step towards the scientific thinking and research. Finally, the mainstream research in the field of quantum theory fully complies with the axiomatic theory [1-3]. Therefore, with some care, a student familiar with the axiomatic, representation-invariant quantum theory can read, and basically follow, every contents based on the formalism of the non-relativistic quantum theory.

The contents of this paper are as follows. In Section 2, we provide the postulates formulations along with the brief remarks, comments and the topics of the research interest. In Section 3 we discuss some prominent benefits
of familiarizing the students with the axiomatic quantum theory. Section 4 is discussion and conclusion, where we provide our proposal for the Curriculum as well as our experience in teaching quantum mechanics, nuclear physics and quantum information.

\bigskip

{\bf 2. The basic postulates}

\bigskip

This section offers a recapitulation of the basic postulates of the standard quantum-mechanical theory [4, 10-13] while  keeping in mind the current state of the art in the field. The explicit formulations of the postulates are compilations or a ''distillate'' of what can be found in the literature. Brief comments following the postulate-formulations cannot possibly exhaust the emphases that should be provided in the full presentation of the contents, e.g. during the lessons. Our aim is to be minimalist, nevertheless fully aware of the unavoidable personal choices and preferences. We state certain postulates that are often used as the work rules for the theory, typically not explicitly formulated as the quantum-mechanical postulates.

	The students should become aware of the phenomenological origin of the quantum mechanical postulates and therefore that the standard postulates may be just one possible basis of the theory. Also, the students should be encouraged to take the proper courses before the start, e.g. of the classical-mechanics Hamilton formalism, basic statistical physics, theory of probability and the basic course in linear algebra--the latter, e.g., a la Vuji\v ci\' c [18], on which basis the Hilbert and the rigged Hilbert space can be smoothly introduced [19]. Equipped with this knowledge, a student will easily follow the contents of quantum mechanics with the benefits emphasized in Section 3 of this paper. Everywhere in this paper we use the standard Dirac notation.

\bigskip

{\bf 2.1 Quantum kinematics}

\bigskip

{\bf Postulate I: Quantum States}. Every state of a quantum system is represented by an element (a vector) of a linear vector space, which is the state space of the system, and vice versa: every element of the vector space is a possible state of the quantum system. Two vectors, $\vert \varphi\rangle$ and $\vert \chi\rangle$, that satisfy the equality

$$
\vert \varphi\rangle = e^{\imath \delta} \vert \chi\rangle \eqno(P.1)
$$
								
\noindent for arbitrary $\delta$, should be regarded the same quantum state.

\begin{table}[h!]
 \caption{Quantum states.}
\centering
    \begin{tabular}{  | p{2cm} | p{13cm} |}
    \hline
    Remarks & The state spaces are different for different kinds of physical systems. It can be finite or infinite dimensional and determined by certain phenomenological rules or constraints, such as e.g. the superselection rules. Building the state space for a system is at the root of doing quantum mechanics, with the free choice of representation. \\ \hline
    Comments   & Linear superposition of quantum states (of the state-space vectors) is a reminiscence of the classical ''superposition of waves'', which may lead to interference. Nevertheless, a quantum state is neither a classical wave nor a classical particle, nor it is uniquely related to the ordinary three-dimensional space. An emphasis should be placed on the fact, that some states need not be in the domain of certain quantum observables, see Postulate II below.\\ \hline
   Research & The set of the available (accessible, or allowed) states for the system can be dynamically redefined, e.g. in quantum decoherence--dynamical superselection rules induced by the system's environment. Whether quantum state is ''realistic'' or merely a source of information (i.e. ''epistemic'') is a long standing problem of vivid current interest. \\
    \hline
 \end{tabular}
\end{table}

{\bf Postulate II: Quantum Observables}. Every variable of a classical system is represented by a Hermitian (self-adjoint) operator on the state space of the system (that is established by Postulate I). And vice versa: every Hermitian operator on the system's quantum state space corresponds to a physical variable (physical quantity) that, in principle, can be physically measured (''observed'').

\begin{table}[h!]
 \caption{Quantum observables.}
\centering
    \begin{tabular}{  | p{2cm} | p{13cm} |}
    \hline
    Remarks & Measurability of a quantum observable assumes that its eigenstates form a complete set, i.e. an orthonormalized basis in the state space. \\ \hline
    Comments   & Measurability of an observable should be taken with care. It may regard operational accessibility (in a lab) or be a subject of additional (e.g. phenomenological) rules.\\ \hline
   Research & The set of the {\it practically} measurable observables may be dynamically determined--e.g. the environment-induced preferred observable(s) in quantum decoherence. \\
    \hline
 \end{tabular}
\end{table}

	In classical mechanics, the basic pair of the {\it variables}--position $\vec r$ and momentum $\vec p$--of a particle determine the particle's {\it state}, $(\vec r, \vec p)$. That is, everything that is needed is at one place: On the one hand, $\vec r$ and $\vec p$ are the physical {\it variables}, i.e. the measurable quantities. On the other hand, {\it as a pair}, those variables define the system's {\it state}—-the ''phase space'' of states that itself is a linear vector space. Every classical state $(\vec r, \vec p)$ {\it uniquely} determines the value of {\it every} possible classical variable $  A = A(\vec r, \vec p)$.

	This elegant classical picture is destroyed by the above quantum postulates. In quantum theory, ''states'' are still elements of a linear vector (state) space, but the observables do not ''constitute'' the quantum states. Quantum states and quantum observables are irreducibly separated.  That is,

\noindent (Q) Knowledge about the system's state does not uniquely determine the values of {\it all} the system's observables. Furthermore, non-commutatitivity of the quantum observables does not even allow existence of  the
(simultaneous) values of all observables.

Bearing (Q) in mind, it seems unavoidable to expect uncertainty, i.e. probabilistic quantum theory. That is, in order to avoid probabilities, we need one-to-one relation between the states and the values of {\it all} observables--as it is the case in classical mechanics.

{\bf Postulate III: Measurement Probabilities}. For a measurement of an observable $\hat A$, with the spectral form
$\hat A = \sum_n a_n \hat P_n + \int_{\alpha}^{\beta} \vert a\rangle a da \langle a\vert$, that is performed on the system in the state $\vert \varphi\rangle$, the probability for the result to fall within some interval
$(c, d)$ reads:

$$
P(\hat A, \vert \varphi\rangle, (c,d))= \langle \varphi \vert \hat P_{(c,d)}(\hat A)\vert \varphi\rangle, \eqno(P.2)
$$

\noindent where $\hat P_{(c,d)}(\hat A)$ is the so-called spectral measure for the interval $(c,d)$ determined by the observable $\hat A$.

\begin{table}[h!]
 \caption{Quantum probabilities.}
\centering
    \begin{tabular}{  | p{2cm} | p{13cm} |}
    \hline
    Remarks & The postulate regards measurement of a single observable and is historically known as the Born's rule. Whether or not some observables can be simultaneously measured requires additional concepts--e.g. ''complete observable'' (equivalently, the complete set of mutually commuting observables). \\ \hline
    Comments   & The measurement result of an observable $\hat A$ is determined by the set of its eigenvalues and the spectral form. Extending the postulate by stating the form of the final state of the object of measurement should be separately performed. To this end, the so-called projective (von Neumann) measurement is the ultimate basis of the "generalized measurements" formalism.
    Noncommutativity of certain observables implies non-existence of the spectral measure, which might be common for the noncommuting observables. \\ \hline
   Research & It is a true challenge to detect the situations in which joint measurements of non-commmuting observables may be possible, at least partly. The so-called quasidistributions (like the Wigner function) may be useful in this regard. Modern statistical (so-called ''device-independent'') approaches to quantum foundations may go even beyond the above presented basic postulates. \\
    \hline
 \end{tabular}
\end{table}

{\bf Postulate IV: Quantization}. Transition from the classical variables to the quantum observables, i.e. quantization of the classical variables of a system, is such that:

(a) The basic set of the independent variables (i.e. the degrees of freedom) becomes a basic set of the  observables. By definition, the degrees-of-freedom-observables commute with each other.
The classically allowed values are assumed to be the eigenvalues of the related quantum observables.

(b) A linear sum of classical variables transforms into the same-form linear sum of the related quantum observables:

\begin{equation} \nonumber
\alpha A + \beta B \to \alpha \hat A + \beta\hat B, \quad \alpha,\beta \in R
\end{equation}

(c) A product of a pair of variables is mapped into the ''symmetrized'' product of the observables:

\begin{equation} \nonumber
AB \to {1\over 2} (\hat A \hat B + \hat B \hat A).
\end{equation}

(d) A Poisson bracket is mapped to a commutator:

\begin{equation} \nonumber
[A,B]_{PB} \to {-\imath \over \hbar} [\hat A, \hat B].
\end{equation}

\noindent with the Planck constant $\hbar$.
	
(e) Transition from the classical to the quantum quantities is continuous.

\begin{table}[h!]
 \caption{Quantization rules.}
\centering
    \begin{tabular}{  | p{2cm} | p{13cm} |}
    \hline
    Remarks & There are alternative quantization schemes. Here presented one is the most common in the nonrelativistic context. \\ \hline
    Comments   & Every classical degree of freedom, $q_i$, is accompanied with its conjugate momentum, $p_i$, $[\hat q_i, \hat p_i]=\imath \hbar$. Mutual independence of the degrees of freedom implies $[\hat q_i, \hat q_j] = 0, [\hat p_i, \hat p_j] = 0, [\hat q_i, \hat p_i]=\imath \hbar\delta_{ij}$, where $\delta_{ij}$ is the ''Kronecker delta''. Due to the point (a), every observable is an analytic function of the basic set of the observables, $\forall \hat A = A(\hat q_i, \hat p_i)$.\\ \hline
   Research & The absence of non-commutativity for classical variables poses a challenge for the transition from the quantum to the classical formalism--a subject of e.g. quantum decoherence and the quantum measurement theory. \\
    \hline
 \end{tabular}
\end{table}

	The following three postulates (V-VII) may not be universally acknowledged.

{\bf Postulate V: Quantum degrees of freedom}. Quantization of a classical degree of freedom, $q_i$, gives the quantum mechanical observable, which (together with its conjugate observable, if such exists) acts on a related Hilbert space $H_i$. The total state space of a system, $H$, is tensor-product of the (''factor'') spaces corresponding to the individual degrees of freedom:

$$
H = \otimes_i H_i. \eqno(P.3)
$$

\begin{table}[h!]
 \caption{Quantum degrees of freedom.}
\centering
   \begin{tabular}{  | p{2cm} | p{13cm} |}
    \hline
    Remarks & The postulate equally regards the classically known, mutually independent, degrees of freedom (such as the Descartes $x,y$ and $z$ coordinates) as well as the phenomenologically defined ''internal'' degrees of freedom, such as the spin, whose components do not mutually commute.\\ \hline
    Comments   &Some internal degrees of freedom, e.g. the spin, do not mutually commute and therefore all act on the same, non-factorizable state space. \\ \hline
   Research &Alternative degrees of freedom can be obtained via the classically-analogous, e.g., linear canonical transformations. Those degrees of freedom define the alternative tensor-factorizations of the system's state space. In general, starting from the set of mutually independent (commuting) degrees of freedom leads to the so-called Tsirelson's problem of whether or not the tensor factorization of the state-space, i.e. eq.(P.3), might have any alternative. \\
    \hline
 \end{tabular}
\end{table}

{\bf Postulate VI: Single systems}. For a quantum ensemble in a state $\vert \varphi\rangle$, every single element of the ensemble should be regarded to be in the same state $\vert \varphi\rangle$.

\begin{table}[h!]
 \caption{Quantum single systems.}
\centering
   \begin{tabular}{  | p{2cm} | p{13cm} |}
    \hline
    Remarks & Postulate III implies a large number of individual acts of a measurement (in accordance with the statistical definition of probability). A set of the individual measurements regards the individual systems that constitute a ''statistical ensemble''. Individual elements of the ensemble must be identically prepared and measured, and must not mutually physically interact. \\ \hline
    Comments   & This is the ultimate basis for introducing the ''mixed'' quantum states, when the observer is not sure in which (''pure'') state the measured single system actually is. Knowing the quantum state uniquely, i.e. with certainty, to be some ''pure state'' $\vert \varphi\rangle$ is the situation of the maximum possible information about the system.\\ \hline
   Research & Description of single systems and their behavior is essential in certain applications (e.g. quantum metrology and the emerging technologies) as well as for the interpretational corpus of the quantum theory. \\
    \hline
 \end{tabular}
\end{table}

{\bf Postulate VII: Composite systems [Nonidentical quantum particles]}. Every composite quantum system consisting of $N$ nonidentical quantum (sub)systems is joined the (total) Hilbert space, $H_{tot}$, that is a tensor product of the Hilbert spaces for the constituting subsystems:

$$
H = \otimes_i H_i. \eqno(P.4)
$$

\begin{table}[h!]
 \caption{Composite quantum systems.}
\centering
    \begin{tabular}{  | p{2cm} | p{13cm} |}
    \hline
    Remarks & Quantum mechanics is not sensitive to the number of the constituent particles of a composite system. That is, the number $N$ in eq.(P.4) may in principle go to infinity. Nevertheless, physically interesting are the finite systems--finite $N$. \\ \hline
    Comments   & Every subsystem of a composite system may itself have whatever degrees of freedom (including the internal ones, such as the spin). For every degree of freedom, arbitrary representation may be chosen. \\ \hline
   Research & Where is the line dividing small (micro, i.e. quantum) and the macro (i.e. many-particle, classical) systems? This is an aspect of the measurement problem, but also of the modern open systems theory (and quantum decoherence), quantum foundations of the thermodynamic relaxation, also of interest in the interpretations of quantum mechanics. \\
    \hline
 \end{tabular}
\end{table}

{\bf Postulate VIII: Identical quantum particles}. (A) A set of $N$ mutually identical particles of the half-integer spin ({\it fermions}), $s=(2n+1)/2,n=1,2,3,…$, are described by quantum states that are {\it antisymmetric} under the particles permutations; (B) A set of N mutually identical particles of the integer spin ({\it bosons}), $s=0,1,2,3,…$, are described by quantum states that are {\it symmetric} under the particles permutations.

\begin{table}[h!]
 \caption{Identical quantum particles.}
\centering
    \begin{tabular}{  | p{2cm} | p{13cm} |}
    \hline
    Remarks & {\it Phenomenology} establishes the superselection rules--independently for the fermions and bosons. This is an instance of the phenomenological suspension of (P.4) that additionally defines the state space, which is abstractly introduced by Postulate I. \\ \hline
    Comments   & Presenting the fermions state by Slater determinant trivially leads to the Pauli Principle, which is phenomenologically known from the atomic physics context. So the Pauli Principle is not really a principle, but a (trivial) corollary of Postulate VIII.\\ \hline
   Research & It is a phenomenological rule that, e.g., electrons belonging to different atoms need not be subject of PP; then the physical situation is described by the postulates I-VII. Building a molecule from a set of atoms typically implies application of PP to {\it all} electrons, i.e. introduces indistinguishability of the molecule's electrons. So, where is the line dividing identical and non-identical fermions in a composite quantum system? If it is not a sharp dividing-line, then ''how well'' is fulfilled the Pauli Principle? \\
    \hline
 \end{tabular}
\end{table}

\bigskip

{\bf 2.2 Quantum dynamics}

\bigskip

{\bf Definition 1}. By {\it isolated} quantum system, it is assumed a system that is not in interaction with any other physical system. An isolated system may be subjected to some external field, which may be time-dependent.

{\bf Definition 2}. By a quantum system's {\it Hamiltonian}, it is assumed a Hermitian observable obtained from the quantization (see Postulate IV) of the classical Hamilton function (energy) of the system.

{\bf Postulate IX: The Schrodinger Law [Quantum determinism]}. Dynamics of an isolated quantum system is a map of the system's state from the initial to the final instant of time, $\vert \varphi(t_{\circ})\rangle \to \vert \varphi(t)\rangle, t \ge t_{\circ}$. This map is a dynamical map that is generated by the system's Hamiltonian so that:

(i) $\alpha \vert \varphi(t_{\circ})\rangle + \beta \vert \chi(t_{\circ})\rangle \to \alpha \vert \varphi(t)\rangle + \beta \vert \chi(t)\rangle $,

(ii)	The state changes deterministically: the state in an instant of time $t_{\circ}$ uniquely determines the state in every later instant of time $t$,

(iii)	$\langle \varphi(t)\vert \varphi(t)\rangle =1, \forall t$,

(iv)	The map $\vert \varphi(t_{\circ})\rangle \to \vert \varphi(t)\rangle$ is continuous in time, i.e. $t$ is a continuous real parameter with the dimension of time.

\begin{table}[h!]
 \caption{Quantum determinism.}
\centering
    \begin{tabular}{  | p{2cm} | p{13cm} |}
    \hline
    Remarks & The concept of isolated quantum system is an {\it idealization}. Practically there is no such thing in a lab. E.g., a decaying/absorbing atom changes the state of the vacuum/surrounding-electromagnetic-field. The systems interacting with the other surrounding systems are {\it not} isolated. \\ \hline
    Comments   & From the postulate it directly follows the unitary (time reversible) dynamics for isolated systems, while introduction of the system's Hamiltonian as the dynamics generator can be performed in non-unique way.
    Unitary character of quantum dynamics of isolated systems is analogous with the classical dynamics in that it establishes deterministic dynamics and {\it unique trajectory in the system's state-space}. This applies also  to the non-conservative systems, which are subject of certain time-dependent external fields (external potentials). Due to Postulate II, dynamics can be independently formulated in the terms of the system's observables--the so-called Heisenberg picture. The quantum dynamical pictures are mutually equivalent and can be combined to give the equivalent, so-called, ''interaction picture''. \\ \hline
   Research & The realistic systems non-trivially interact with other systems. The general rules for such, so-called {\it open} systems are not yet known--open quantum systems (and decoherence) theory. \\
    \hline
 \end{tabular}
\end{table}

\bigskip

{\bf 3. Usefulness of the axiomatic approach}

\bigskip

Axiomatic quantum mechanics provides the shortest and most efficient path to familiarizing with the basics and universal use of quantum mechanics. It provides the basic methodological core and approach to every scientifically-useful presentation of the quantum-mechanical theory as well as its upgrades towards the diverse applications. Below, we emphasize some specific benefits of making the students familiar with the axiomatic quantum mechanics.
Representation-invariance is also a precursor for the quantum field theory, notably for the free Dirac field.

I) It is possible to clearly distinguish quantum kinematics from quantum dynamics--very much like the standard attitude and benefits known from the classical mechanics. Certain specific benefits are emphasized below.

II) From Postulates I and IX, it is clear that the ''wave function'', $\varphi(\vec r, t)$, is nothing but one out of the plenty possible, the so-called {\it position-representation} of the quantum state $\vert \varphi(t)\rangle$; in Dirac notation, $\varphi(\vec r,t) = \langle \vec r\vert\varphi(t)\rangle$.  Representation should be avoided as long as it is possible since the basic formulas can all be written in the {\it representation-independent} Dirac form. E.g. the measurement probability, see (P.2) for the notation:

\begin{equation}
P(\hat A, \vert \varphi(t)\rangle, a_n) = \langle\varphi(t)\vert \hat P_n\vert \varphi(t)\rangle.
\end{equation}
					
\noindent  The choice of the representation should be made such that the calculation of (1) is made easier. The same applies to the choice of the dynamical picture; e.g., in the Heisenberg picture, eq.(1) reads:

\begin{equation}
P(\hat A(t), \vert \varphi(0)\rangle, a_n) = \langle\varphi(0)\vert \hat P_n(t)\vert \varphi(0)\rangle,
\end{equation}

\noindent That is, all the basic expressions practically directly follow from the Postulates--no need for memorizing.

III) Non-cummutativity of the position and momentum operators implies both, nonexistence of the common representation, i.e. of the representation of the form $\varphi(\vec r, \vec p, t)$, as well as nonexistence of the common spectral measure, which might lead to the {\it exact} simultaneous measurements of those observables.

IV) Postulates I and II point out and emphasize the substantial divorce of ''quantum'' from ''classical'' in that, as distinct from the classical systems, in quantum theory there is the [quantum] information limit. That is, every pure state carries the maximum information about the system that can be acquired by measurement. Nevertheless, for every pure state $\vert \varphi\rangle$ exist certain quantum observables for which the state is not an eigenstate. Hence the non-unique values of such observables for the system in the state $\vert \varphi\rangle$. In contrast to this, ''pure'' classical states (e.g. the points in the classical phase space) give unique value for every possible physical variable of the system. For this reason, it is said that there are {\it no dispersion-free quantum ensembles}. This is the essence of ''quantum uncertainty'' that is so often misused (or even abused) in presentation of the quantum theory.

V) Quantum uncertainty may be quantified by the standard deviation, $\Delta \hat A = \sqrt{\langle \hat A^2\rangle - \langle \hat A\rangle^2}$, of an observable $\hat A$; the mean value of the observable $\hat A$ in the state $\vert \varphi\rangle$, $\langle \hat A\rangle = \langle \varphi\vert \hat A\vert \varphi\rangle$. The uncertainty relations due to Robertson [20] is a direct consequence of Postulates I and II, i.e. a theorem of quantum mechanics:

\begin{equation}
\Delta \hat A \cdot \Delta \hat B \ge {1\over 2} \vert \langle [\hat A, \hat B] \rangle\vert,
\end{equation}

\noindent which is  often incorrectly interpreted. The clear distinction of quantum kinematics and dynamics, Section 2, distinguishes eq.(3) as a {\it kinematical} result that does not regard quantum measurements in the terms of dynamical chronology or ''measurement disturbance''; eq.(3) also applies for $\vert \varphi(t)\rangle$ in an arbitrary but {\it fixed}, single instant of time $t$. By definition, the standard deviations regard the {\it statistical ensembles}, not the single quantum systems, that clearly opens the door for avoiding misinterpretations of eq.(3). Borrowing from [1] [and retaining our notation]:

\noindent
''[...] {\it if we prepare a large number of quantum systems in identical states, $\vert \varphi\rangle$, and then perform measurements of $\hat A$ on some of those systems, and of $\hat B$ in others, then the standard deviation
$\Delta\hat A$ of the $\hat A$ results times the standard deviation $\Delta\hat B$ of the results for $\hat B$ will satisfy the inequality} (3).''

\noindent That is, eq.(3) neither regards a single quantum system nor does it describe dynamical ''measurement disturbance'' on the system. Dynamics of a single measured system and the thereof disturbances are subject of the other kinds of ''uncertainty relations'' that investigate the Heisenberg's intuition on quantum measurement and are of interest e.g. in quantum information and quantum metrology. The two kinds of quantum uncertainty relations  regard the mutually exclusive physical scenarios.

VI) Postulate II assumes that every observable has the complete set of eigenstates (eigenvectors) thus directly leading to the concept of simultaneous measurability in quantum mechanics. Every pair of observables, $\hat A$ and $\hat B$, which are some functions of a {\it complete} observable (all eigenvalues are non-degenerate), $\hat E$, are {\it simultaneously measured} (i.e. appearing with the unique values) {\it whenever the observable} $\hat E$ {\it is measured}. However, this implies commutativity $[{\hat A, \hat B}] = 0$ and emphasizes the lack of the rules for simultaneous measurability of the mutually non-commuting observables. Non-commuting observables can be simultaneously measured on an ensemble in a state, which is a common eigenstate for the observables. Simultaneous measurability of the commuting observables does not imply unique values for a pair of commuting observables in every possible state; this can be easily seen by considering a state that is an eigenstate of the one, but not of the other observable.

VII) Postulate VI directly distinguishes the following typical assertion in doing quantum mechanics: ''If a quantum system is in a state $\vert \varphi\rangle$ ...''. The immediate question, at least justified by imprecision of the laboratory procedures, reads: ''What if we are not sure in which state the single system actually is?'' This subjective lack of information leads to unavoidable use of quantum ''mixed'' states (often called ''proper mixtures'') that are presented by statistical operators (''density matrices''). Therefore the main interpretation of the proper mixed states: every single system is in one state out of the set of the possible states, not in a coherent superposition of any of those states. Nevertheless, bearing in mind the mathematical non-uniqueness of the form of the statistical operators, this interpretation should be carefully used: it certainly works if we know, in advance, which quantum states are mixed.

VIII) Postulate VIII distinguishes the fundamental role of phenomenology in formulation of non-relativistic quantum theory. In conjunction with Postulate I and Postulate VII, it emphasizes a need for careful reading and interpretation of the very basic postulates about the quantum-system’s states. A careful reader will immediately realize that Postulate VIII requires non-tensor-factorization of the state space of identical particles (bosons or fermions)--i.e. {\it non-validity} of eq.(P.4). Hence the lesson for teaching quantum mechanics: {\it quantum mechanical postulates should not be separately presented or discussed}. Finally, Postulate VIII is the ultimate basis of the famous Pauli Principle. Actually, quantum state $\vert \psi\rangle$ of a set of identical fermions can be written in the form of the Slater determinant, e.g. for three fermions:

\begin{equation}
\vert \psi\rangle_{123} = {}\sqrt{{1\over 3!}}  \left| \begin{array}{ccc}
\vert \varphi\rangle_1 & \vert \chi\rangle_1 & \vert \phi\rangle_1 \\
\vert \varphi\rangle_2 & \vert \chi\rangle_2 & \vert \phi\rangle_2 \\
\vert \varphi\rangle_3 & \vert \chi\rangle_3 & \vert \phi\rangle_3 \end{array} \right|
\end{equation}							

\noindent Setting, e.g., $\varphi = \chi$ in the determinant makes the two columns of the determinant mutually equal and hence directly leads to $\vert \psi\rangle_{123} = 0$, i.e. to impossible quantum state for a set of three identical fermions. Physical meaning of this is historically known as the Pauli Principle: no pair of mutually identical fermions can take the same quantum state. Therefore, Pauli Principle is a (trivial) implication of Postulate VIII, not really a quantum-mechanical principle.

IX) Bearing in mind atomic phenomenology, Definition 1 and Postulate IX distinguish the physical atoms as {\it open}, not isolated quantum systems. Phenomenology reveals emission and/or absorption of photons by atoms that is historically described by the rules for ''quantum jumps'' of the atom, practically without  considerations of the electromagnetic field. However, the inevitable change of the electromagnetic field (EMF, which becomes the quantum vacuum in the absence of photons) during the atomic emission/absorption processes reveals that the electromagnetic field is a {\it dynamical system} that cannot be ignored in any model or interpretation. Instead of the sole ''atom'' system, phenomenology implies the composite system ''atom+EMF''. If the atom were isolated, it would not be able to emit or absorb photons. The proof for this claim is rather easy. The standard atomic ''stationary states'' are eigenstates of the atomic Hamiltonian. Unitary dynamics established by Postulate IX then implies temporal persistence of every eigenstate $\vert \varphi_n\rangle$ of the atomic Hamiltonian, $\hat H$ ($\hat H \vert \varphi_n\rangle = E_n \vert \varphi_n\rangle$). That is, [bearing in mind Postulate I, i.e. eq.(P.1)], the unitary dynamics due to Postulate IX:

\begin{equation}
\hat U(t) \vert \varphi_n\rangle = e^{-\imath \hat H t/\hbar} \vert \varphi_n\rangle = e^{-\imath E_n t/\hbar} \vert \varphi_n\rangle = \vert \varphi_n\rangle
\end{equation}

\noindent so the ''quantum jumps'', $\vert \varphi_m\rangle \to \vert \varphi_n\rangle, m\neq n$, are {\it not possible for an isolated atom}. Quantum modeling of the composite system ''atom+EMF'' may be a matter of taste [2, 5, 6, 8] but the fact that every atom is an open system is a direct consequence of Postulate IX in conjunction with the atomic phenomenology.

X) Following the postulates, it is straightforward  to "decipher" the physical meaning of certain formulas not known to the student, with the basic information about the underlying model. E.g. the standard expression of the electric quadruple moment of the atomic nucleus  that regards the nucleus as a rotating rigid body, not as a point-like particle, e.g. eq.(15.35) in Ref. [21]:

\begin{equation}
Q = 2\int \psi^{\ast} r^2 P_2(\cos\theta) \psi d{\bf r}.
\end{equation}

\noindent The {\it classical} model of "rigid body" assumes the external center-of-mass ($CM$)  and the Euler angles for the nucleus rotational degrees of freedom. In this model, the internal spatial degrees of freedom are defined for the bulk of the rigid body--the position-vector $\vec r$ that is accompanied by the mass and the electric charge densities (the total positive charge being $Ze$). Ignoring the $CM$ dynamics can be performed by placing the reference frame into the $CM$ system. Then remain the internal spatial and the external rotational degrees of freedom of the body. For this model, Postulate IV and Postulate V imply the related tensor-factorization of the quantum nucleus' state space,
$H_{internal}\otimes H_{rotational}$, which should be extended by the protons' spin state space which can be collectively denoted by $H_{spin}$:

\begin{equation}
H_{nucleus} = H_{internal}\otimes H_{rotational} \otimes H_{spin} \equiv H_{internal}\otimes H_{J};
\end{equation}

\noindent the "factor"-space $H_J$ regards the total angular momentum of the nucleus, i.e. the observable $\hat {\vec J} = \hat {\vec L} + \hat {\vec S}$, where $\hat {\vec L}$ stands for the nucleus' total angular momentum and $\hat {\vec S}$ the nucleus' total spin observable. Then the classical definition of the electric quadruple moment, $Q = 3 z^2 - r^2$, due to Postulate V, directly gives the observable $\hat Q = 3\hat z^2 - \hat r^2$ for the internal spatial observable $\hat {\vec r}$; rigorously, bearing eq.(7) in mind, $\hat{\vec r} \equiv \hat{\vec r}\otimes \hat I_{J}$. For the Descartes degrees of freedom, $\hat{\vec r} = \{\hat x, \hat y,\hat z\}$, the tensor-factorization $H_{intrinsic} = H_x\otimes H_y\otimes H_z$ directly follows due to Postulate V. Transition to the spherical degrees of freedom (suggested by $r$ and $\theta$ in eq.(6)), due to Postulate V, gives rise to re-factorization: $H_{internal} = H_r \otimes H_{\Omega}$, where the nucleus' angular momentum can be shown to read $\hat I_r \otimes \hat{\vec L}$, i.e. acting only on the spatial-angle factor space $H_{\Omega}$. Therefore, in the position-representation, the standard expression $z = r\cos\theta$ directly gives: $3 z^2 - r^2 = r^2 (3\cos^2\theta - 1) \equiv 2r^2 P_2(\cos\theta)$, where $P_2$ stands for the Legendre polynomial, $P_2(x) = (3x^2 -1)/2$. Then the origin and the meaning of eq.(6) becomes obvious: $Q=\langle \psi\vert \hat Q\vert\psi\rangle$, while bearing in mind that $\psi$ appearing in eq.(6) assumes the position-representation (spherical coordinates) for the tensor factorization eq.(7).

\noindent The alternative model regards the atomic nucleus as a set of the protons and neutrons modelled as the point-like particles, i.e. the set $\{p_i, n_j, i=1,2,...,Z, j=1,2,...,A-Z\}$, where $Z$ and $A$ stand for the atomic and the atomic-mass numbers. Then the basic degrees of freedom $\{\vec r_{pi}, \vec r_{nj}, i=1,2,...Z, j=1,2,...A-Z\}$ constitute the classical model of the nucleus. Following Postulate IV and Postulate V, those degrees of freedom act on the related factor-spaces of the total Hilbert state space:

\begin{equation}
H = \otimes_{i=1}^{Z} H_{pi} \otimes_{j=1}^{A-Z} H_{nj}.
\end{equation}

\noindent Of course, every nucleon's state space (for every index $i$ and $j$ in eq.(8)) is  tensor product of the orbital and the spin factor spaces, $H^{(orbital)} \otimes H^{(spin)}$. Adopting the spherical coordinates for every nucleon separately as described above introduces the individual nucleon's state space factorization: $H_r\otimes H_{\Omega}\otimes H^{(spin)}$. Placing this and grouping the angular- and the spin-factor-spaces gives:

\begin{equation}
H = \otimes_{i=1}^{ Z} H_{pri} \otimes_{j=1}^{A-Z} H_{nrj} \otimes_{i=1}^{Z} H_{p\Omega i}\otimes_{j=1}^{A-Z} H_{n\Omega j} \otimes_{i=1}^{Z} H^{(spin)}_{pi} \otimes_{j=1}^{A-Z} H^{(spin)}_{nj} = \otimes_{i=1}^{ Z} H_{pri} \otimes_{j=1}^{A-Z} H_{nrj} \otimes H_J,
\end{equation}

\noindent where the $H_J$ space regards the nucleus' total angular moment as defied above. Ignoring the neutrons gives rise to the electric quadruple in the position-representation [22]:

\begin{equation}
Q_p = \sum_{i=1}^Z (3 z_{pi}^2 - r_{pi}^2)
\end{equation}

\noindent as a discrete form of eq.(6), while the factorization eq.(9) incorporates eq.(7).

XI) Now it is ready to build on the postulates of Section 2 in order to introduce: (A) Axiomatic theory of "mixed states" as distinguished in Table VI, towards the axiomatic open quantum systems theory [8]; (B) From Postulates VI and IX it directly follow that interaction in composite systems leads to the formation of entangled quantum states, on which basis it can be introduced the other non-classical kinds of quantum correlations as the possible resources for quantum information processing [1, 23]. E.g. for the {\it dominant}  interaction of the form $\hat A_1 \otimes \hat B_2$, with the indices referring to the interacting systems, $1$ and $2$, and with the eigenvalues $a_i$ and eigenstates $\vert i\rangle_1$ of $\hat A_1$, it dynamically follows entangled state for the $1+2$ composite system:

\begin{equation}
e^{-\imath \hat A_1\otimes\hat B_2 t/\hbar} \vert \phi\rangle_1 \otimes \vert \chi\rangle_2  = e^{-\imath \hat A_1\otimes\hat B_2 t/\hbar} \sum_{i} c_i \vert i\rangle_1 \otimes \vert \chi\rangle_2 = \sum_i c_i \vert i\rangle_1 \otimes \left(e^{-\imath a_i\hat B_2 t/\hbar} \vert \chi\rangle_2\right);
\end{equation}

\noindent (C) In analogy with our emphasis on the quantum kinematics and dynamics, the kinematic and dynamic symmetries can be introduced without a special preparation of students for the new contents. The group of the particles permutations is a basis for both, the second quantization formalism as well as for the  quantum field theory.

XII) Extending Postulate III with the final state of the object of measurement can be performed in non-unique way. Nevertheless, it can be shown that the so-called projective (von Neumann) measurements, together with Postulate IX, represent the ultimate basis for all known kinds of measurements, such as the "generalzied measurement" and the so-called POVM measurements [1].

XIII) Re-examination of the basic postulates is the core of the current quantum-mechanical research that includes the efforts of re-formulation of the basics of the theory. Notably, the need for Postulate III emphasizes insufficiency of Postulate IX for describing the process of quantum measurement. Just like the atoms commented in VII, the object of quantum measurement is not an isolated but open quantum system. Description of quantum measurements is a part of the ongoing efforts to describe dynamics and behavior of the general open quantum systems [2, 8], still with significant contributions from the interpretational corpus of the quantum mechanical theory.

\bigskip

{\bf 4. Discussion and conclusions}

\bigskip

The desired "visualizations" and "explanations" of the quantum mechanical formalism should better be left to the specialized applications and interpretations of quantum mechanics. Often, they produce the puzzles and make the theory "mysterious"
before becoming useful for certain {\it limited purposes}. For example, the idea of the electron orbiting around the proton in the hydrogen atom may be useful in certain limited contexts of the atomic physics but appears
to come to a flat contradiction with the fact [7] that the electron and the proton are quantum-mechanically {\it entangled} with each other. Equally unreliable are the statements regarding the fate of a {\it single} object of quantum measurement, especially in attempts of "explaining" quantum uncertainty (of any kind).

In our opinion, the main goal of teaching the axiomatic quantum mechanics should be to emphasize its {\it basic, methodological character} that enables the upgrades towards the specialized courses in non-relativistic quantum physics and some of the prominent current scientific research in physics and emerging technologies. We believe that such a course can be properly presented on about 150 pages. To this end, we recognize  Chapter 2 of Nielsen and Chuang's [1], which is announced in the book's Preface as follows:

\noindent ''{\it Aside from classes on quantum computation and quantum information, there is another
way we hope the book will be used, which is as the text for an introductory class in quantum
mechanics for physics students}.''

The needs of quantum information and computation mainly regard the finite-dimensional quantum systems (qubits and their realizations). Nevertheless, if equipped with the formalism needed for the continuous (''continuous variable'') systems, the Nielsen-Chuang’s intro (currently around 60 pages) in quantum mechanics could indeed be used as a basic course of quantum mechanics on the undergraduate/graduate level. 	Therefore we conclude that we are still missing a proper textbook, which would present the axiomatic quantum theory in a concise yet sufficient form for the first encounter of physics students with the quantum mechanical theory.

From Sections 2 and 3 we construct the following {\it sketch} of the Curriculum for the introductory course of quantum mechanics for the physics students:

-- Quantum kinematics: Postulates I through VIII.

-- Building the functional state space for the continuous systems.

-- Quantum dynamics: Postulate IX (leading to the Schr\" odinger equations).

-- The general theory of angular momentum.

-- The Stern-Gerlach experiment and the theory of the spin-1/2.

-- Solutions of the Schr\" odinger equation (conservative systems, bound states): simple one-dimensional models, harmonic oscillator and the hydrogen atom.

-- Perturbation theory.

-- Generalized quantum measurement.

-- Non-relativistic quantum symmetries (kinematical: the extended Galilei group; dynamical: symmetry group of the system's Hamiltonian).

Hence the Lego-dice-like upgrades toward the modern topics, such as:

-- Basics of the quantum scattering theory (continuous spectrum of the Hamiltonian--related to the so-called scattering states in the Hilbert state space).

-- Composite quantum systems: Quantum entanglement (kinematical aspect: the Schmidt canonical form; dynamical aspect: interactions in  composite systems).

-- Non-classical correlations and their measures.

-- Axiomatic formalism of the "mixed" quantum states.

-- Quantum subsystems: Non-unitary dynamics, "improper" mixed states, basic concepts of the open-systems theory.

-- Selected chapters of quantum interpretations (physical nature of "quantum state"; quantum measurement and the transition from quantum to classical; hidden variables) etc.

-- Second quantization formalism.

\noindent Certain combinations of these upgrades with the basic Curriculum may be useful  for studies of some related fields and applications in modern science and technology.

Our experience in teaching quantum mechanics, nuclear physics and quantum information emphasizes usefulness of the axiomatic quantum theory. E.g. from the total of 94 students, 57\% of them have successfully passed the exam. If we do not account the students that have not regularly attended the lectures, the percent goes to more than 76! The average mark is between C and D (numerically 7.43). Similarly, teaching nuclear physics is much easier with the underlying basic course of quantum mechanics. Of the total of 60 students, around 40\% have passed the exam. However, accounting only the students who regularly attended the lessons raises the percent to approximately 73! The average value is approximately $C$ (numerically: 8.3). We find those scores encouraging: the students properly prepared and active during the lessons did not have any serious problems in adopting the
topics of the basic course in quantum mechanics as well as its application in nuclear physics.
So, in our teaching practice, we are convinced of the famous statement of Boltzmann:

\noindent''{\it Nothing is more practical than a good theory}.''

So we conclude: the needs for application of quantum theory should be clearly separated from the basic formalism on the graduate/undergraduate level of physics studies. Introductory course of quantum mechanics can be formulated in the axiomatic form with significant benefits for students regarding both, more specialized applications as well as familiarizing with the current scientific research.

\begin{acknowledgments}

Work on this paper is financially supported by Ministry of Science Serbia, grant 171028.

\end{acknowledgments}

\end{document}